\documentclass[fleqn,usenatbib]{mnras}

\usepackage{newtxtext,newtxmath}
\usepackage[T1]{fontenc}

\DeclareRobustCommand{\VAN}[3]{#2}
\let\VANthebibliography\thebibliography
\def\thebibliography{\DeclareRobustCommand{\VAN}[3]{##3}\VANthebibliography}

\usepackage{graphicx}	% Including figure files
\usepackage{amsmath}	% Advanced maths commands
\usepackage{todonotes}

\title[IRAS04125 Unbound]{Disc-planet misalignment from an unstable triple system: IRAS04125}

\author[Rebecca Nealon et al.]{Rebecca Nealon,$^{1,2}$\thanks{E-mail: rebecca.nealon@warwick.ac.uk}
Jeremy L. Smallwood,$^{3}$
Hossam Aly,$^{4}$
Andrew J. Winter,$^{5, 6}$
Cristiano Longarini,$^{7}$ \newauthor
Nicolás Cuello,$^{8}$
Dimitri Veras$^{1,2,9}$ and
Richard Alexander${^{10}}$
\\
$^{1}$Centre for Exoplanets and Habitability, University of Warwick, Coventry CV4 7AL, UK\\
$^{2}$Department of Physics, University of Warwick, Coventry CV4 7AL, UK\\
$^{3}$Homer L. Dodge Department of Physics and Astronomy, The University of Oklahoma, Norman, OK 73019, USA\\
$^{4}$Faculty of Aerospace Engineering, Delft University of Technology, Kluyverweg 1, 2629 HS Delft, The Netherlands\\
$^{5}$Max-Planck Institute for Astronomy (MPIA), Königstuhl 17, 69117 Heidelberg, Germany \\
$^{6}$Universit{\'e} C{\^o}te d'Azur, Observatoire de la C{\^o}te d'Azur, CNRS, Laboratoire Lagrange, 06300 Nice, France \\
$^{7}$Institute of Astronomy, University of Cambridge, Madingley Road, Cambridge, CB3 0HA, UK \\
$^{8}$Univ. Grenoble Alpes, CNRS, IPAG, 38000 Grenoble, France \\
$^{9}$Centre for Space Domain Awareness, University of Warwick, Coventry CV4 7AL, UK \\
$^{10}$School of Physics \& Astronomy, University of Leicester, University Road, Leicester, LE1 7RH, UK
}
\date{Accepted XXX. Received YYY; in original form ZZZ}

\pubyear{\the\year{}}

\begin{document}
\label{firstpage}
\pagerange{\pageref{firstpage}--\pageref{lastpage}}
\maketitle

\begin{abstract}
The IRAS$01425$+$2902$ wide binary system was recently reported to have both a young planet and a puzzling geometric arrangement, where the planet and binary both orbit edge-on, but misaligned by $60 \degr$ to the circumprimary disc. This is the youngest transiting planet yet to be detected but its misalignment to the disc is difficult to explain. In this paper we explore the dissolution of an unstable triple system as a potential mechanism to produce this system. We simulate the effects of an ejection interaction in models using a highly inclined, retrograde flyby centred on the primary star of IRAS$01425$. The escaping star of $\sim 0.35$ M$_{\odot}$ inclines both the disc and binary orbits such that they have a relative misalignment of $\gtrsim60 \degr$, as inferred from observations. The planet orbit also becomes inclined relative to the disc, and our interpretation predicts that the binary should have a highly eccentric orbit ($e\gtrsim0.5$ from our simulations). We additionally demonstrate that despite the high relative misalignment of the disc it is unlikely to be vulnerable to von Zeipel-Kozai-Lidov oscillations.
\end{abstract}


%%%%%%%%%%%%%%%%%%%%%%%%%%%%%%%%%%%%%%%%%%%%%%%%%%

%%%%%%%%%%%%%%%%% BODY OF PAPER %%%%%%%%%%%%%%%%%%
\section{Introduction}
The mysteries presented by IRAS$04125$+$2902$ (hereafter IRAS$04125$) are typical of misaligned or warped discs: it is a wide binary system ($\gtrsim500$ au) with a transition disc extending from $20-60$ au in the dust around the primary star \citep{Espaillat:2015ve}. Recent observations from \citet{Barber:2024te} discovered a $ 10.7 \, R_\oplus$ radius planet orbiting the primary star, with a mass of $<0.3$ M$_{\rm J}$ and an orbital period of $8.83$ days. This planet was important for two reasons; first, it was the youngest transiting planet detected to date (with a system age of $\sim 3$ Myr) and second, the planet orbital plane was found to be strongly inclined with respect to the disc orientation, although the position angle is not constrained. The disc is misaligned by $60 \degr$ to the plane of the outer binary orbit, the latter being edge-on (as for the transiting planet). Given the apparent edge-on orbits of both binary and planet, \citet{Barber:2024te} conclude that the most likely arrangement is the planet-binary orbits are aligned, while both misaligned with respect to the disc.  If this is indeed the case, then how to produce a disc with such an extreme misalignment in an otherwise co-planar system is an intriguing puzzle. In this work, we explore whether dynamical interaction between members of a multiple system is a plausible way to produce such a system.

IRAS$04125$ is a member of the Taurus star forming region. The typical stellar density averaged over the whole volume of Taurus is too low for random encounters within the continuum disc radius ($\lesssim 60$~au) over the disc lifetime to be plausible \citep[where we define an encounter to be when two unbound stars come within 1000~au,][]{Winter:2024tm}. During  the hierarchical star formation process, interactions between multiple systems are expected to be common, however these interactions are typically confined to the first few $10^5$~yr after formation \citep{Bate:2018ls,Delgado-Donate:2004iw,Offner:2023as}. Nonetheless, within the surviving sub-groups in Taurus \citep{Joncour:2018tn}, binary-single scattering events (or equivalently the decay of chaotic triples) can continue to produce occasional disc-penetrating encounters up to the $\sim 3$~Myr age of Taurus. Possible observational examples include UX Tau \citep{Menard:2020} and RW Aurigae \citep{Dai:2015yr}, both of which underwent disc-disrupting encounters $\sim 10^3$~years ago, and the HV Tau triple system, which appears to have ejected DO Tau in a violent disc-disc encounter $\sim 0.1$~Myr ago \citep{Winter:2018hv}. 

Given the serendipitously uncovered examples of such recent encounters, this would strongly suggest that star-disc encounters in multiple systems are relatively common occurrences. Indeed, in their full dynamical model of the Taurus region, \citet{Winter:2024tm} estimate ten severely disc-truncating encounters (removing $>10$~percent of the disc mass) per Myr throughout the whole region by the present age of Taurus ($\sim 1-3$~Myr). In this model, all such encounters occur during interactions between three or more stars. The fact that IRAS$04125$ remains a binary system, and that the binary companion is misaligned with respect to the dusty disc \citep{Barber:2024te}, is further circumstantial evidence for a star-disc encounter originating from the dissolution of a chaotic triple system scenario. 

A highly inclined object perturbed by an outer companion also undergoes von Zeipel–Lidov–Kozai (vZKL) oscillations, where its orbital inclination and eccentricity are periodically exchanged \citep{vonZeipel:1910nh,Kozai:1962gq,Lidov:1962gq}. These oscillations are recovered in simulations - not only in tilted satellites but also in misaligned protoplanetary discs \citep[e.g.,][]{Martin:2014aa,Smallwood:2023kj}. The critical tilt required to induce vZKL oscillations in a test particle is about $39^\circ$, similar to that for a protoplanetary disc \citep[e.g.,][]{Martin:2014aa}.  
The disc observed in IRAS$04125$ appears to meet this criterion, with a relative misalignment of $60\degr$. It would then be expected that the orbital arrangement of IRAS$04125$ will drive vZKL oscillations, but in a disc these oscillations are damped towards the critical tilt due to viscous torques. Importantly, vZKL oscillations move away from the maximum inclination and towards maximum eccentricity --- in other words, vZKL oscillations are not able to increase the relative misalignment towards $60 \degr$ as the initial misalignment must have been greater than or equal to this. Given that the damping to the critical angle probably occurs on time-scales comparable to the disc lifetime \citep[e.g. viscous time-scale][]{Martin:2014bb,Ceppi:22024vc}, this would suggest that the disc reached a more extreme misalignment ($
\gtrsim 60^\circ$) with the binary orbit relatively recently.  

In this letter we present a potential origin of the enigmatic arrangement of IRAS$01425$ using the dissolution of a chaotic triple system. We approximate the dynamics of this triple star interaction using a close-in flyby encounter and show that a low mass ejector could drive the required misalignments between the planet, disc and binary orbital planes.

\section{Modelling a chaotic triple system}
Finding initial conditions to model a chaotic multiple scattering event including a hydrodynamic disc is a computationally intractable problem. In the following, we therefore model the system as the primary star, a binary companion that remains in the system, and what we hereafter call a `flyby' star, that can be understood to be the ejected member in a chaotic triple star interaction. This simplifies the search for initial conditions, and is practically equivalent to modelling the end stages of the chaotic ejection. We caution that this reduces our study to a proof of concept, with the goal of demonstrating if a triple system can yield large relative misalignments as a result of a chaotic dynamical encounter.

Previous work by \citet{Pfalzner:2009pj,Nealon:2018bt} shows that by itself, only a flyby of several times the primary mass would be capable of inclining the disc relative to the original orbital plane. But the largest star in Taurus is AB Aur with $\sim 2.4$ M$_{\odot}$ \citep{DeWarf:2004oj} and there are only a handful of stars with $> 1$ M$_{\odot}$ \citep{Luhman:2004bu}, making it impossible to motivate such a massive flyby. In this work, we thus appeal to a flyby perturbing both the disc and binary to achieve the required relative inclination between the two and seek to understand if it can be also be used to misalign the planet orbit. As we shall show, it is possible to use a flyby to generate the required misalignments suggesting that a single flyby could explain the geometry of IRAS$04125$. However, our models also show that this requires relatively a large perturber mass. A single flyby is thus unlikely to be a complete interpretation as stars this massive are uncommon in the Taurus region.

To model the flyby interaction with the disc we choose to use the 3D smoothed particle hydrodynamics (SPH) code \textsc{Phantom} \citep{Phantom}. SPH is well suited to investigate flyby simulations \citep[e.g.][]{Clarke:1993bv} and \textsc{Phantom} in particular has extensively been used to examine their evolution \citep[e.g.][]{Cuello:2019bd, Nealon:2019oh, Smallwood:2024yq}. We adopt typical protoplanetary disc parameters as a best approximation of the disc \citep[see e.g.][]{Nealon:2018ic}. We also assume an aspect ratio (between the scale height $H$ and cylindrical radius $R$) $H/R = 0.05$ at $20$ au and a \citet{shakura_sunyaev} viscosity $\alpha=0.001$ such that the disc is safely in the `wave-like' regime where warp propagation is governed through pressure effects \citep{Papaloizou:1995pn}, as appropriate to protoplanetary discs.

To represent the IRAS$01425$ system we use a disc mass of $M_{\rm d} = 2 \times 10^{-3}$ M$_{\odot}$ \citep{Barber:2024te} with $R_{\rm in} = 20$ au and $R_{\rm out} = 60$ au \citep[but note that the inner radius could be smaller,][]{Espaillat:2015ve}. In gas-dust calculations of flybys, \citet{Cuello:2019bd} showed that the dust is truncated more effectively than the gas during a flyby due to radial drift effects, thus we can reasonably assume that the outer edge of the gas disc is $>60$ au. We consider $R_{\rm out} = 75$ and $100$ au in our gas only calculations here.

We use $M_1=0.7$ M$_{\odot}$ and $M_2 = 0.17$ M$_{\odot}$ as the primary and secondary masses of the binary \citep{Barber:2024te}. We simplify the problem significantly by assuming a circular orbit for this binary and use the projected separation of $a_{\rm b}=635$ au. We also assume this binary starts in the $x$-$y$ plane and use this as our initial reference. All of our simulations are run to at least two binary orbits, $T_{\rm b}$ (corresponding to $\sim44$ orbits at $75$~au or $\sim28$ at $100$~au). Here we note an important and consequential limitation of our simulations as our choice of $a_{\rm b}$ is a lower estimate. Our simulations are somewhat degenerate to this choice however; we could have used a larger $a_{\rm b}$, $R_{\rm out}$ and flyby mass to achieve the same relative inclination.

We set the flyby as in \citet{Cuello:2019bd} on a parabolic orbit initialised at $20 R_{\rm min}$ so that it is started outside the orbit of the outer binary for any choice of $R_{\rm min}$ (the distance of minimum approach). For the flyby star we investigate masses of $M_{\rm f} = 0.14, 0.35, 0.70, 1.05$ and $1.4$ M$_{\odot}$
(corresponding to $0.2, 0.5, 1.0, 1.5$ and $2\times$ the primary star mass). For all simulations the inclination is set to $i = 135\degr$ to maximise the disc tilt and minimise disc destruction \citep{Xiang-Gruess:2016bq,Nealon:2019oh,Cuello:2023nn} such that the flyby is retrograde to the gas in the disc. The inclination $i$ of the flyby's orbit is set by assuming a co-planar flyby and then rotating the position and velocity vectors through the inclination angle around the $y$ axis. Here we consider an extra orbital element in the flyby initial conditions in order to quantify the impact of the orbital phase of the binary. This was achieved by keeping the binary orbit consistent and rotating the flyby orbit through an approach angle $\beta$ around the $z$ axis. We consider $12$ different $\beta$ spread evenly between $\beta=0\degr$ and $\beta=360\degr$.

\begin{figure}
    \centering
    \includegraphics[width=\linewidth]{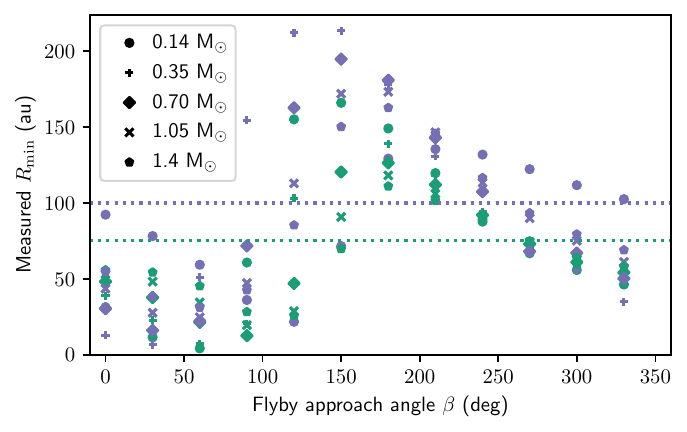}
    \caption{The minimum approach of the flyby to the primary star as measured from our simulations with $R_{\rm out} = 75$ au in green and $R_{\rm out} = 100$ au in purple. The outer edge of the disc is indicated with a dotted line and the symbols to different flyby masses. Between $\beta=120 \degr$ and $\beta=240 \degr$ the flyby does not interact strongly with the disc. Rather, in these cases the relative misalignment is due to the perturbation of the outer binary star.}
    \label{fig:minimum_approach}
\end{figure}

In order to ensure that the flyby is timed just right for a grazing encounter (where $R_{\rm min} = R_{\rm out}$), we initialise the binary such that the primary star is always located at the same absolute position at the instant of pericentre passage. The centre of the flyby passage is then set to occur around this location and meets the primary star at the exact moment to make the encounter grazing. To achieve this we vary the true anomaly for each simulation but note that because $e=0$ initially for the binary orbit, this is equivalent to changing the argument of periapsis. We calculate the initial true anomaly of the outer binary orbit $f_i$ by comparing the time until the flyby occurs $T_f$ to the orbital time-scale of the binary $T_b$. The flyby time-scale is calculated using Barker's equation as in Appendix~A of \citet{Cuello:2019bd}\footnote{We note a typo in Equation A7, where $2 r_{\rm peri}^3$ should read $(2 r_{\rm peri})^3$, as is used in \textsc{Phantom}.}. The true anomaly is then set to $f_i = -T_f / T_b \times 360$ in degrees. As in Figure~\ref{fig:minimum_approach}, our method means that the average minimum separation between the stars across the different angles of approach for the flyby $\beta$ averages to the chosen $R_{\rm min}$, but as the flyby encounter occurs 3-body dynamics takes over causing some variation for each simulation.

In a subset of our simulations we add a planet with the inferred properties of IRAS~$04125+2902$~b. Since a dynamic encounter can tighten a planet's orbit \citep{Heggie:1975be}, it could be the case that the planet's current location ($0.074$ au) is its final resting place and before the interaction the planet was located further out. We thus consider two orbital radii; one corresponding to the inner edge of the dust cavity \citep[$20$ au,][]{Espaillat:2015ve} and one to the currently observed location \citep[$0.074$ au,][]{Barber:2024te}. The primary and outer binary stars have accretion radii (the radius within which gas is accreted) of $R_{\rm acc} = 10$ au while the flyby has $R_{\rm acc} = 1$ au. In the simulations in Section~\ref{section:withplanet} this means that the planet may be within the accretion radius of the primary star and as such does not interact with the gas. The planet is set to have an accretion radius of $R_{\rm acc} = 0.25R_{\rm Hill}$ where $R_{\rm Hill}$ is the radius of the Hill sphere \citep[e.g.][]{Nealon:2018ic}.

\begin{figure}
    \centering
    \includegraphics[width=\linewidth]{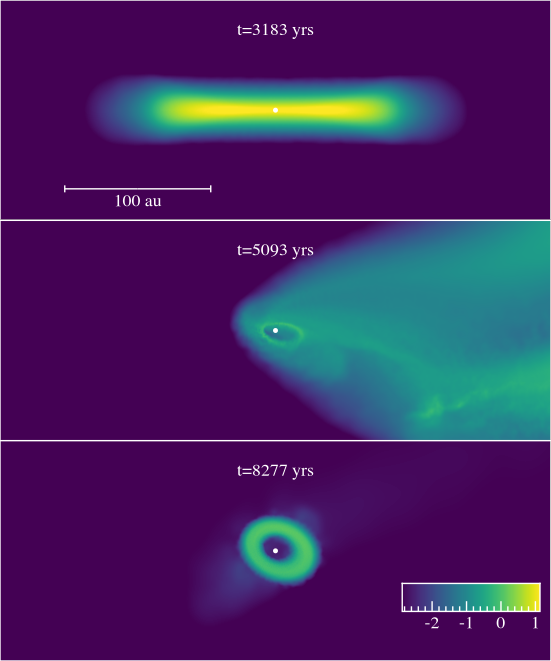}
    \caption{Explaining the misaligned disc in IRAS$04125$ using a flyby encounter: here the flyby has $M_{\rm f}=0.35$ M$_{\odot}$, $R_{\rm out}=75$ au and $\beta=60 \degr$. The encounter is disruptive and the remaining disc is torqued to more than $60 \degr$. The panels are all shown in the $x$-$z$ plane to emphasise the disc rotation with a scale $100$ au and the stars shown with white dots. The upper panel shows before the encounter, the middle panel during and the lower panel long after the encounter. The colour represents log of the column density in cgs}.
    \label{fig:rendered_example}
\end{figure}

\section{Simulations resembling a stellar ejection}
\label{section:main_results}
Figure~\ref{fig:rendered_example} illustrates the indicative evolution of one of our simulations with a column density rendering. In each simulation, as the outer binary orbits the primary the flyby moves towards the primary star and has an encounter with the disc. This perturbation drives the disc to change its orientation and maintain it as the flyby leaves. The outer binary orbit is also strongly affected by its interaction with the flyby, changing the orientation of its orbit, becoming eccentric and on occasion becoming unbound from the system entirely. Because the disc is in the `wave-like' regime \citep{Papaloizou:1995pn} and radially narrow, the rapid warp propagation means that we find no evidence of disc breaking during any of our simulations despite the large torque that is applied.

\begin{figure}
    \centering
    \includegraphics[width=\linewidth]{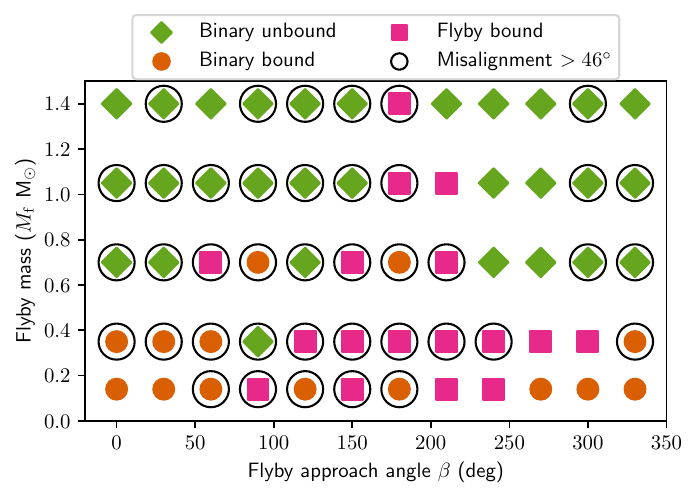}
    \caption{Flyby encounters that could be consistent with IRAS~$04125$ are shown with orange symbols circled in black. Results are shown as a function of the flyby approach angle $\beta$ against the mass of the flyby $M_{\rm f}$ for $R_{\rm out}=75$ au (the $R_{\rm out}=100$ au are qualitatively similar, with most successes around $M_{\rm f} =0.35$ M$_{\odot}$). Orange circles indicate when the binary remains bound and flyby remains unbound, green diamonds when the binary becomes unbound, pink squares when the flyby becomes bound.}
    \label{fig:simulations_summary}
\end{figure}

\subsection{Binary-disc misalignment}
\label{section:outer_flyby_only}
We now explore different flyby masses $M_{\rm f}$, angles of approach $\beta$ and outer radius $R_{\rm out}$. For each simulation there are three metrics needed to establish whether the escaping star has altered the binary system in a way that would be consistent with IRAS$04125$. Figure~\ref{fig:simulations_summary} shows the summary of our results for \text{an outer radius} of $75$ au (noting that the results for an outer radius of $100$~au are qualitatively similar). In Figure~\ref{fig:simulations_summary} we display the most favourable simulations with orange symbols circled in black.

First, the relative inclination between the binary orbit and the mass weighted average of the disc must be consistent with observations. While the observed value is $\sim 60 \degr$, taking into account uncertainties it could be as low as $\sim 46 \degr$ \citep{Barber:2024te}. We thus consider anything with $\gtrsim 46 \degr$ to be `successful', indicated with black circles in Figure~\ref{fig:simulations_summary}. We find misalignments between $46 \degr$ to $\sim 170 \degr$ are possible between the disc and binary. We set this lower bound because a relative misalignment of more than $46\degr$ can drive vZKL oscillations that could then evolve the relative misalignment to exactly the observed value. Lower flyby masses do not incline the system strongly enough but higher masses are far more likely to unbind it.

Of our simulations that achieve a disc-binary misalignment of $> 46\deg$, we caution that when $120 \degr < \beta < 210 \degr$, this is mostly due to orbital evolution of the binary and \emph{not} torquing of the disc. Figure~\ref{fig:minimum_approach} explains this; for $120 \degr < \beta < 210 \degr$ the measured minimum approach of the flyby is larger than the outer radius of the disc and so it does not drive a meaningful change in the disc orientation. In this scenario, the planet and disc maintain their shared orbital alignment which is inconsistent with \citet{Barber:2024te}. We thus consider `success' to be in simulations where the encounter is disc-penetrating or disc-grazing (noting that all are binary-penetrating), i.e. $\beta < 120 \degr$ or $\beta > 210\degr$.

Second, the binary star must remain bound to the system and the perturbing star must remain unbound - either of these would be inconsistent with observations. We test this by measuring the energy of the secondary and flyby assuming they are in orbit around the primary. In Figure~\ref{fig:simulations_summary} this is indicated with orange circles when the binary remains bound and the flyby unbound, green diamonds for where the binary becomes unbound and pink squares when the flyby becomes bound. In line with our expectations from \citet{Sigurdsson:1993om}, we find that the binary is more likely to remain bound for the lower mass flybys (circles), with the vast majority of simulations above $m_{\rm f}=0.70$ M$_{\odot}$ unbinding from the flyby. Although not shown for our full parameter suite, our results also robustly predict that the binary should become eccentric as a result of the flyby encounter, with $e>0.5$ for all of our successful combinations --- noting that this could be consistent with observations since the binary orbit is poorly constrained.

Third, the location of the outer edge of the gas disc is crucial. Realistically, we expect that the outer radius of the disc will be truncated by both the interaction with the ejected star (our `flyby') and pericentre passage of the binary. However, in our simulations we have underestimated the binary orbit and it becomes eccentric during the interaction, shrinking the pericentre distance further - any outer radius measured after subsequent binary passages will be too small and misleading. We thus measure the radius after the flyby passage but \emph{before} the binary star does a pericentre passage. In practice, we look for the moment when the radius of the flyby is greater than that of the binary, i.e. when the flyby is no longer significantly interacting with the orbits of the primary and binary. While we cannot exclude any of our simulations due to the subsequent interactions with the binary, we expect that the pericentre distance of the actual binary orbit must be greater than $60$ au.

To quantify the outer radius of the gas disc after the encounter, we isolate the material bound to the primary and identify the radius that contains $68 \%$ of this mass, yielding outer radii averaging between $30-120$ au after the encounter.

Our results demonstrate that the relative misalignment in IRAS$01425$ can be driven by a low mass flyby, similar to the encounter expected from the dissolution of a chaotic triple system. Because the outer radius in our models does not include the long term effects from the tidal truncation of the binary, they represent the largest values we might expect from these limited simulations. However, because of our lower limit assumption for the semi-major axis of the binary they are also smaller than we might expect. In any case, our moderate underestimation of the outer disc radius is only in mild tension given uncertainties inherent in comparing simulated gas to continuum (mm dust) radius. While typically exceeding dust radii by a factor $\sim 2$ \citep{Ansdell:2018vi, Sanchis:2021gd}, gaseous outer radii based on optically thick CO line emission probe out to very low outer surface density \citep[for a discussion of associated caveats, see][]{Miotello:2023do}. This ejected star is most likely $\sim 0.35$ M$_{\odot}$ and had a disc-penetrating, retrograde encounter with the disc of IRAS$01425$. Our results show that remaining binary must have an eccentric orbit and suggests that the outer radius of the disc (and thus minimum approach of the perturber) was larger than $100$ au.

\subsection{Planet-disc misalignment}
\label{section:withplanet}

\begin{figure}
    \centering
    \includegraphics[width=\linewidth]{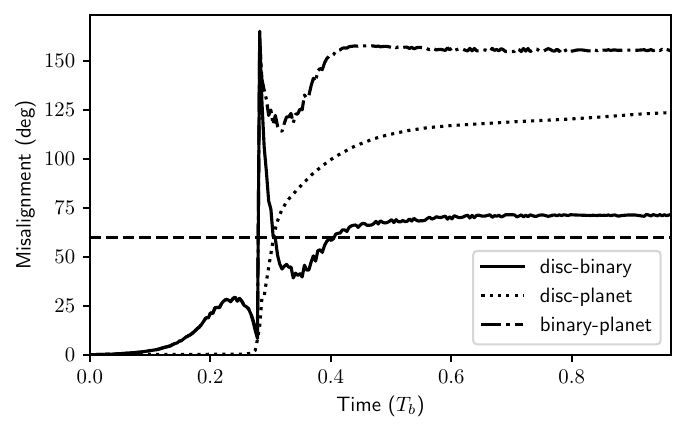}
    \caption{The disc-planet and disc-binary inclination throughout the ejection encounter for the simulation shown in Figure~\ref{fig:rendered_example}. Solid represents the disc-binary misalignment (as in the right panel of Figure~\ref{fig:simulations_summary}), dotted shows the disc-planet misalignment and dash-dot the binary-planet. The horizontal dashed line represents the observed disc-binary misalignment \citep{Barber:2024te}.}
    \label{fig:planet_properties}
\end{figure}

We now consider the evolution of the planet in this system with two additional simulations. For the simulation shown in Figure~\ref{fig:rendered_example} ($M_{\rm f} = 0.35$ M$_{\odot}$, $\beta=60 \degr$ and $R_{\rm out} = 75$ au) we add a planet at $0.074$ au and separately at $20$ au. Here we present the results for the $0.074$ au planet but note the results are qualitatively similar for the planet starting at $20$ au. Figure~\ref{fig:planet_properties} shows the inclination of the disc relative to both the planet and binary orbits. We find that the planet remains bound to the primary star throughout the disruption with little change to its radius, oscillating between $0.08-0.12$ au. The planet orbit becomes strongly inclined to the disc ($\sim 124 \degr$) and binary ($\sim 150 \degr$) after the flyby interaction.  We repeat this process for the `success' simulations in Figure~\ref{fig:simulations_summary}, finding planet-disc misalignments between $\sim 6 \degr$ to $\sim124 \degr$. From our limited sample, our results also show that the planet-binary inclination is greater than $70 \degr$ for all of the `success' simulations that we ran, with a couple consistent with retrograde alignment (i.e. inclined by $\sim 180 \degr$ relative to binary). \citet{Barber:2024te} suggest but cannot fully constrain the planet-binary alignment to within $\sim 10 \degr$. In light of our results, if the planet was subsequently observed to be aligned to the binary it is likely to be retrograde.

\subsection{Feasibility of companion ejection}
\begin{figure}
    \centering
    \includegraphics[width=\linewidth]{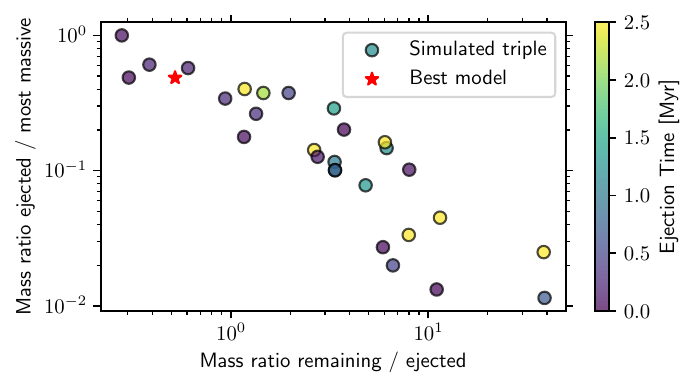}
    \caption{The mass ratios of the components of triple systems that ejected one of its members in one of the simulations of the entire Taurus star forming region by \citet{Winter:2024tm}. The points are coloured by the time at which one of the components is ejected. We mark as a red star the adopted ratios for the model favoured by the parameter space exploration in this work. }
    \label{fig:triples_decay}
\end{figure}

Generally, we expect dynamical ejection to statistically favour specific mass ratios for the ejected stars \citep[e.g.][]{Manwadkar:2021te}. We can ask whether the decay of a chaotic triple system that ejects a star of mass $0.35\, M_\odot$ from an IRAS$04125$-like system is statistically likely in the Taurus region - noting that only a handful of these lead to the geometry we are looking for (Figure~\ref{fig:simulations_summary}). To answer this question, we take an N-body simulation of \citet{Winter:2024tm} and extract all stellar triple systems that undergo dynamical decay in their model of Taurus (which reproduces observed structure metrics). We then show in Figure~\ref{fig:triples_decay} the mass ratios of the components compared to those of our `best-fit' model. During the lifetime of Taurus, we find 28 examples of ejection of a triple system, 16 within $1$~Myr. Interestingly, the locus of favoured mass ratios is in excellent agreement with the best model from the hydrodynamic simulations. This not only indicates that one such encounter could feasibly occur in Taurus, but the fact that the independently constrained escaper mass ratio is statistically favoured dynamically circumstantially supports the chaotic triple scenario. This is further strengthened by the `best-fit' model comfortably corresponding to a region of the parameter space with an ejection timescale of $<1$~Myr, in agreement with the young age of the system.

\section{von Zeipel-Kozai-Lidov oscillations}
\label{section:ZKL}
Section~\ref{section:main_results} demonstrates that $46 \degr$ misalignment of the disc in IRAS$01425$ can be explained by an ejection style event. However, it could also be the result of von Zeipel-Kozai-Lidov (vZKL) oscillations or even these two mechanisms acting in concert. The vZKL oscillation of a disc may be suppressed by the gas pressure given the criterion: 
\begin{equation}
    S = 0.36 \bigg( \frac{a_{\rm b}}{3R_{\rm out}}\bigg)^3 
    \frac{M_1}{M_2} \bigg(\frac{H(R_{\rm out})}{0.1R_{\rm out}}\bigg)^2 \lesssim 1, 
    \label{eq::S_value}
\end{equation}
where $H$ is the disc scale-height \citep{Zanazzi2017}. Physically, $S^{-1}$ quantifies the strength of the tidal torque (per unit mass) exerted on the outer disc by the external companion, relative to the torque generated by the gas pressure. For a given value of $R_{\rm out}$, we calculate the corresponding $H$ using a simple irradiated flaring disc model \citep{Chiang1997}, assuming the luminosity of the central star in IRAS 04125. 

\begin{figure}
    \centering
    \includegraphics[width=0.45\textwidth]{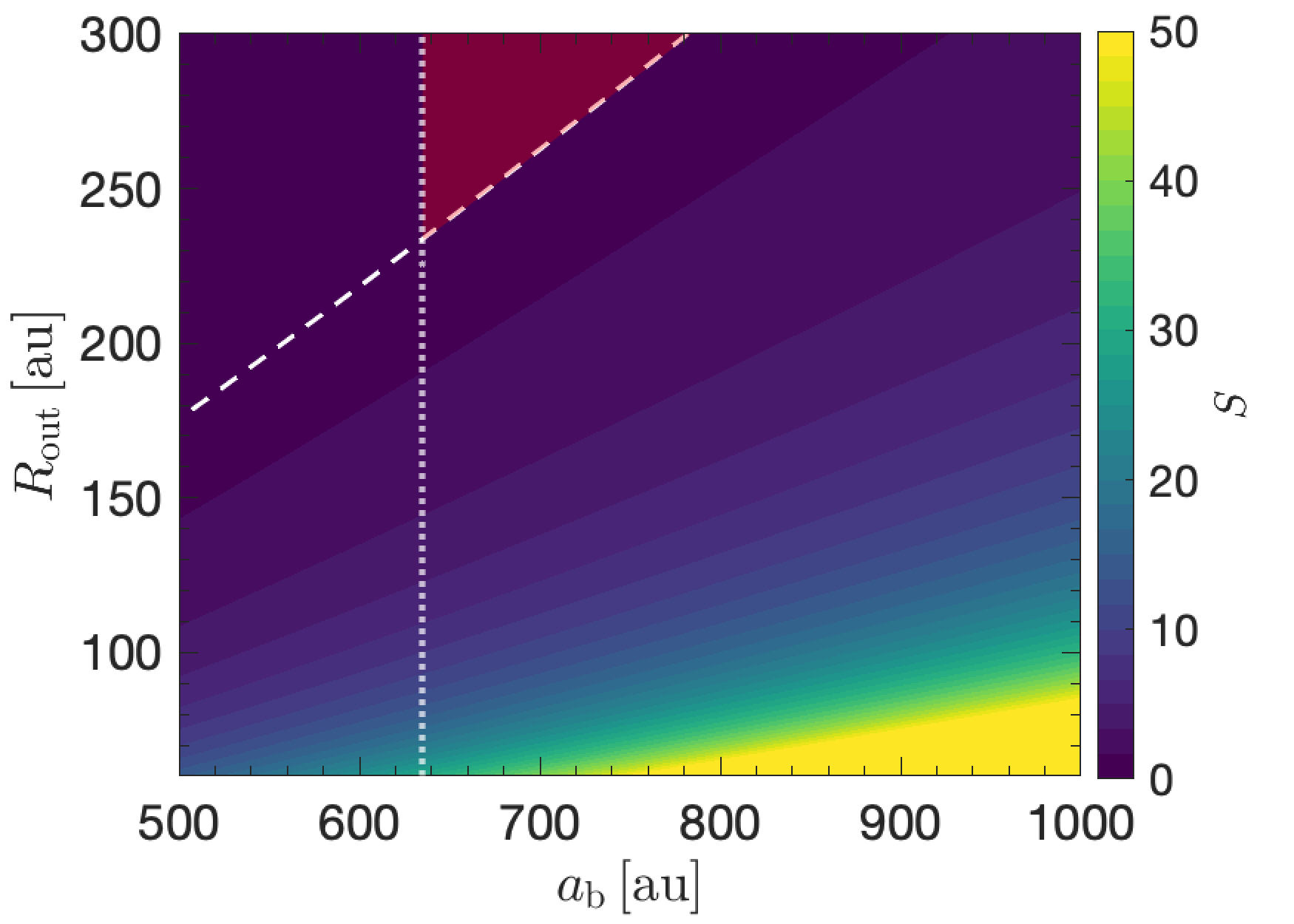}
    \caption{The $S$-parameter from Eq.~(\ref{eq::S_value}) is shown as a function of the binary separation, $a_{\rm b}$, and the outer edge of the gaseous disc, $R_{\rm out}$. The dashed curve represents the points where $S \lesssim 1$. The dotted line corresponds to the lower limit of the projected separation of IRAS\,04125, $635\, {\rm au}$. The red-shaded region indicates the combinations of $a_{\rm b}$ and $R_{\rm out}$ for which the disc is susceptible to the vZLK instability.
    }
    \label{fig::KL}
\end{figure}

Figure~\ref{fig::KL} illustrates the $S$ criterion for inciting the vZLK instability as a function of the binary separation, $a_{\rm b}$, and the outer edge of the gaseous disc, $R_{\rm out}$. The red-shaded region represents the combinations of $a_{\rm b}$ and $R_{\rm out}$ for which the vZLK instability can operate. Given the low mass of the outer companion, the radial extent of the disc must be sufficiently large for the instability to occur. We find that the minimum radial extent of the disc required for the vZLK instability to be active is $R_{\rm out} \sim 230\, \rm au$ (assuming $a_{\rm b}=635$ au). Consequently, a larger disc radius is necessary to allow the vZLK instability to operate, otherwise it is suppressed by the disc gas pressure. As the outer edge of the gas disc is typically a factor of a few larger than the outer edge of the mm dust disc \citep{Ansdell:2018vi,Sanchis:2021gd}, this scenario is still possible but strictly requires that the gas disc be quite extended.

\section{Discussion and Conclusions}
We investigated the dynamics of the IRAS01425 system that hosts a planet, disc and outer star inclined to one another. We have demonstrated that this enigmatic system is probably formed as a result of a low mass encounter, consistent with the dissolution of a chaotic triple system with an ejector mass of $\sim0.35-0.70$ M$_{\odot}$. Our interpretation is favourable because it can naturally explain all puzzling aspects of the currently observed geometry and is motivated by the expected dynamical decay of triple systems in Taurus \citep{Winter:2024tm}. While the final binary orbits from our best fitting simulations underestimate the projected binary orbit (essentially by design), our results do suggest that the binary will be highly eccentric. The observed geometry could also be formed exclusively by vZKL oscillations if the gas disc extends out to $\gtrsim230$ au.

\cite{Barber:2024te} indicate that the inner planet and outer binary might be aligned, although they acknowledge the possibility that they may both be edge-on with different position angles. In this paper, we start our analysis by assuming a primordial alignment between the planet, disc, and outer binary. We show that a third body encounter may be able to reproduce the observed disc-binary misalignment. However, Figure~\ref{fig:planet_properties} shows that these successful encounters induce a planet-binary misalignment. Indeed, our experiments here show that a three body interaction that causes a misalignment between the disc and binary does so by perturbing both inclinations, and therefore will not leave the planet and binary with their original aligned configuration.

If the planet and binary are instead misaligned, one can consider another scenario where the binary is originally misaligned to the disc - indeed, this is the motivation for vZKL oscillations in Section~\ref{section:ZKL}. In addition to the possible vZKL oscillations, the binary could then exert a precession-inducing gravitational torque on the disc. The planet forms in the inclined and precessing disc, acquiring a different precession frequency either after its formation \citep{Nealon:2020vg}, or during the formation process as dust decouples from the gas \citep{Aly:2020ds}. This would ultimately drive a planet orbit misaligned to both the disc and outer binary.

We thus have three possibilities presented by our results:
\begin{itemize}
    \item \textit{The planet and binary orbit are misaligned, and the gas disc outer radius $\lesssim 230$ au}: In this case, vZKL cannot be active and a three body interaction is more probable to cause the observed geometry.
    \item \textit{The planet and binary orbit are misaligned, and the gas disc outer radius $\gtrsim 230$ au}: Then vZKL can produce the observed misalignment with or without the triple perturbation. 
    \item \textit{The planet and binary orbit are aligned}: This is the most challenging case, and would rule out vZKL or a chaotic ejection as the cause of the initial disc misalignment. From our models, we do not find cases where the outer binary is finally aligned with the planet orbit. Such a scenario may be possible, but highly unlikely; in this case another solution would be required, perhaps such as late stage infall.
\end{itemize}

The first of these options is strongly favoured by both our observations and the outcomes from the N-body calculations of \citet{Winter:2024tm}. While our results demonstrate that the arrangement of IRAS$04125$ could arise from a dynamical encounter, our limited knowledge of the semi-major axis and the initial distribution of multiple stellar systems in Taurus precludes a more quantitative analysis. More generally, our findings highlight the role of dynamical interactions in shaping misaligned planetary architectures in multiple star systems. As exoplanet surveys continue to reveal such systems, improved constraints on planet-disc-binary alignments will be key to distinguishing formation pathways and refining our understanding of planetary evolution in complex stellar environments.  

\section*{Acknowledgements}
The authors would like to warmly thank Álvaro Ribas, Grant Kennedy, Vatsal Panwar and Alex Mustill for helpful initial discussions and the referee for constructive comments. RN acknowledges funding from UKRI/EPSRC through a Stephen Hawking Fellowship (EP/T017287/1). JLS acknowledges funding from the Dodge Family Prize Fellowship in Astrophysics, the Vice President of Research and Partnerships of the University of Oklahoma and the Data Institute for Societal Challenges.  HA acknowledges funding from the European Research Council (ERC) under the European Union’s Horizon 2020 research and innovation programme grant agreement No 101054502, AJW under the Marie Skłodowska-Curie grant agreement No 101104656 and NC under grant agreement No. 101042275 (project Stellar-MADE). CL acknowledges funding from the Science \& Technology Facilities Council (STFC) through Consolidated Grant ST/W000997/1 and RA through Consolidated Grant ST/W000857/1. This work was performed using Avon, the HPC clusters at the University of Warwick.

%%%%%%%%%%%%%%%%%%%%%%%%%%%%%%%%%%%%%%%%%%%%%%%%%%
\section*{Data Availability}
The data for Figures~\ref{fig:rendered_example}~and~\ref{fig:planet_properties} are available at 10.5281/zenodo.15170751. All the data underlying this article will be shared on request to the corresponding author. The software used to create and visualise the simulations are publicly available:\\
\textsc{Phantom} \citep{Phantom}:\\
https://github.com/danieljprice/phantom\\
\textsc{splash} \citep{Price:2007kx}:\\
https://github.com/danieljprice/splash\\
\textsc{sarracen} \citep{Sarracen}:\\
https://github.com/ttricco/sarracen



\bibliographystyle{mnras}
\bibliography{overleaf} % if your bibtex file is called example.bib

%\begin{thebibliography}{99}
%\end{thebibliography}

%%%%%%%%%%%%%%%%%%%%%%%%%%%%%%%%%%%%%%%%%%%%%%%%%%

%%%%%%%%%%%%%%%%% APPENDICES %%%%%%%%%%%%%%%%%%%%%

%\appendix

%%%%%%%%%%%%%%%%%%%%%%%%%%%%%%%%%%%%%%%%%%%%%%%%%%

% Don't change these lines
\bsp	% typesetting comment
\label{lastpage}
\end{document}